\documentclass[10pt,onecolumn,showpacs,amsmath,amssymb,aps,nofootinbib,floatfix,preprintnumbers,notitlepage,superscriptaddress]{revtex4-2}
\RequirePackage[colorlinks=true
,urlcolor=blue
,anchorcolor=blue
,citecolor=blue
,filecolor=blue
,linkcolor=blue
,menucolor=blue
,linktocpage=true
,pdfproducer=medialab
,pdfa=true
]{hyperref}
\usepackage{amsmath,amssymb,amsthm,amsfonts}

\usepackage{graphicx,tabularx}
\usepackage{color}
\usepackage{mathrsfs}
\usepackage{multirow}
\usepackage{booktabs}
\usepackage{comment}
\usepackage{enumitem}
\usepackage{cleveref}
\usepackage{slashed}
\usepackage[normalem]{ulem}
\usepackage{scalerel}
\usepackage{wrapfig}
\usepackage{ulem}
\usepackage{cancel}
\usepackage{xcolor}
\usepackage[T1]{fontenc}
\usepackage[utf8]{inputenc}
\usepackage{orcidlink}
\newcommand{\ket}[1]{  \left| #1 \right> }
\newcommand{\bra}[1]{  \left< #1 \right| }
\newcommand{\be}{\begin{eqnarray}}
 \newcommand{\eqn}[1]{Eq.\,(\ref{#1})}

\newcommand{\order}[1]{ \mathcal{O} \left( #1 \right) }

\newcommand{\ee}{\end{eqnarray}}

\newcommand{\ave}[1]{\left\langle #1 \right\rangle}

\newcommand{\eqcomma}{\phantom{AA},\phantom{AA}}

 \usepackage[colorlinks=true, linkcolor=blue, citecolor=blue, urlcolor=blue]{hyperref}
\begin{document}

\title{Hydrodynamical Initial State in Small Systems From the Phase-Space Entropy}

\author{Gabriel Rabelo-Soares \,\orcidlink{0000-0003-3843-4747}}
\email[Corresponding author: ]{grsoares@ifi.unicamp.br}
\affiliation{Universidade Estadual de Campinas - Instituto de Física Gleb Wataghin\\
Rua Sérgio Buarque de Holanda, 777\\
 CEP 13083-859 - Campinas - São Paulo - Brazil}

\author{Gojko Vujanovic \,\orcidlink{0000-0001-5397-6662}}
\email[Corresponding author: ]{gojko.vujanovic@uregina.ca}
\affiliation{Department of Physics, University of Regina, Regina, Saskatchewan S4S 0A2, Canada}

\author{Giorgio Torrieri \,\orcidlink{0000-0002-0611-766X}}
\email[Corresponding author: ]{torrieri@ifi.unicamp.br}
\affiliation{Universidade Estadual de Campinas - Instituto de Física Gleb Wataghin\\
Rua Sérgio Buarque de Holanda, 777\\
 CEP 13083-859 - Campinas - São Paulo - Brazil}


\begin{abstract}
The experimental observation of collective behavior in proton-proton and proton-nucleus collisions poses a fundamental theoretical question regarding the proper characterization of the initial state underlying hydrodynamic evolution. While relativistic hydrodynamics requires an initial condition characterized by an entropy current, corresponding to a maximally mixed state, the microscopic description of the nucleon is based on inherently quantum objects that probes projections of pure states. We show that the appropriate matching between hadronic structure and classical hydrodynamics emerges from the coarse-graining of the partonic Wigner distribution, leading naturally to a Husimi distribution and its associated Wehrl entropy. This entropy provides a semiclassical, positive-definite measure of the density of accessible microstates at a given resolution scale, and therefore constitutes the appropriate quantity to characterize entropy deposition in small collision systems.
\end{abstract}

\maketitle



\section{Introduction \label{secintro}}

For over two decades, the ultrarelativistic heavy-ion program, with LHC and RHIC, has systematically investigated the properties of the quark–gluon plasma (QGP), a phase of quantum chromodynamics (QCD) in which the fundamental constituents of matter are no longer confined within hadrons (such as protons and neutrons). Theoretical modeling \cite{gale_hydrodynamic_2013,romatschke_relativistic_2019,shen_recent_2020} of this exotic phase is consistent with the statement that the QGP behaves as a fluid \cite{heinz_collective_2013}.

Remarkable progress has been achieved on the theoretical modeling of heavy-ion collisions. The current state-of-the-art model of heavy-ion collisions \cite{shen_iebe-vishnu_2016,putschke_jetscape_2019,nijs_bayesian_2021} includes a general framework consisting of initial-state modeling, followed by a pre-equilibrium dynamics and subsequent hydrodynamic evolution, whose continuous degress of freedom are ultimately converted into discrete hadrons further simulated using hadronic transport.

Although the heavy-ion program was originally designed to study nucleus-nucleus (\textit{AA}) collisions, advances in experimental capabilities have allowed the exploration of experimental signatures of collectivity in small collision systems \cite{nagle,atlas_collaboration_measurement_2018,gajdosova_probing_2021}. In heavy-ion (\textit{AA}) collisions, relativistic hydrodynamics is employed to investigate the collective properties of the quark–gluon plasma. From the observed signatures of collectivity in small systems \cite{nagle,atlas_collaboration_measurement_2018}, relativistic hydrodynamics has also been applied to small collision systems. However, this carries several challenges. How to properly characterize the initial conditions in small systems is one of the main issues in the field. 

In this work, we shall explore this question under the ansatz of ``the fastest possible thermalization'' allowed by fundamental principles such as causality and quantum mechanics. Thus, we disconsider any ``kinetic'' entropy production which would be associated with a relaxation/thermalization timescale, and consider local decoherence to be the dominant mode of production of (entanglement) entropy, as hypothesized in \cite{kharzeev}  Such entropy is associated with just a coarse-graining scale, a non-additive local quantity that can be fixed from the total density of degrees of freedom, inferrable from the final multiplicity and the volume.  Once this scale is set, entropy becomes additive, allowing for the projectile, target and any subsystem to be examined separately, ineach unit of area of transverse space (a similar idea was implementd using the ``thin film decoherence approach in \cite{mullerfilm}). We implement the ideas above  by linking the entropy framework (like Shannon and Wehrl entropies) to the initial state description of \textit{pp} and \textit{pA} collisions. This link becomes important owing to the entropy density deposition of such collisions being on the order of the nucleon size. In this situation, quantum fluctuations can no longer be neglected, and the problem reduces to understanding how a quantum system thermalizes (or better yet hydrodynamizes) into a classical system. 

As this work introduces a new formalism for the construction of hydrodynamic initial conditions, we begin by reviewing the relevant theoretical ingredients and establishing the conceptual framework.
The paper is therefore organized as follows. In Section \ref{GPD}, we provide a brief overview of the light-cone wave function formalism and its connection to partonic distributions. We then introduce the Wigner distribution, both in quantum mechanics and in its QCD generalization as well as its coarse-grained version: the Husimi distribution.
In Section \ref{sec:entropy_formalism_for_high_energy_physics}, we discuss the entropy formalism relevant for phase-space descriptions, focusing on the Wehrl entropy as a semiclassical measure of information content. 
Section \ref{initialhydro} reviews the general framework of relativistic hydrodynamics emphasizing how the entropy is encoded into the theory, including  the ideal, viscous and a fluctuating approach appropriate for small systems.
In Section \ref{modelmain}, we demonstrate how the entropy content of the initial state can be constructed from the Wigner distribution through coarse-graining. This procedure naturally leads to a measure of decoherence relevant for hydrodynamization.
Finally, Section \ref{discussion} summarizes the main results, discusses possible phenomenological implications, and outlines how the framework proposed herein can be tested in current and future experiments.

\section{Parton Distribution Functions \label{GPD}}

The foundations of light-cone quantization trace back to Dirac’s classification of relativistic dynamics \cite{Dirac}. In a relativistic setting, the choice of the time parameter is not unique. By analyzing the stability subgroups of the Poincaré group, one finds three natural choices for time, leading to Dirac’s instant, point, and front forms of dynamics. The latter provides the basis for light-cone dynamics. These formulations differ in how the generators of the Poincaré group are separated into kinematical and dynamical ones (see \cite{brodsky} for a detailed review).
In high-energy physics, the front form is particularly useful. In this case, dynamics is defined on the light-cone surface $x^{+} \equiv t + z = 0$, which reduces the number of dynamical generators compared to the instant form. As a result, this framework allows for a simpler and frame-independent description of relativistic wave functions. In this framework, the hadron wave function can be expressed as
\begin{equation}
\label{psidef}
    | \Psi \rangle = \sum_{n = 1}^{\infty} \int d \Phi_n \Psi_n(x_i, \mathbf{k}_{\perp i}) \prod^{n}_{i} a^{\dagger}_i (x_i, \mathbf{k}_{\perp i})|0 \rangle,
\end{equation}
where $d \Phi_n$ is the phase-space  differential, $a^{\dagger}_i (x_i, \mathbf{k}_{\perp i})$ is the parton creation operators acting on the vacuum state $|0 \rangle$ (it creates a parton \textit{i} with longitudinal momentum fraction $x_i$ and transverse momentum $\mathbf{k}_i$), and $\Psi_n(x_i, \mathbf{k}_{\perp i})$ represents the light-cone wave functions of the $n$-parton Fock states. The hadron wave function is a pure state, once the Fock states form a complete basis, the hadron state can be expressed as a superposition $| \Psi \rangle  = \sum_{n} \alpha_n  | n \rangle $, hence its density matrix can be written as
\begin{equation}
    \hat{\rho} =  | \Psi \rangle  \langle \Psi | = \sum_{n,n'} \alpha_n \alpha^{*}_{n'}  | n \rangle  \langle n' |.
    \label{densitymatrix}
\end{equation}
By definition, this state have zero von Neumann entropy \cite{Shannon:1948dpw}. The full quantum information of the hadron, encoded in its wave function, can be equivalently represented in phase-space through quantum Wigner distributions that takes the form
\begin{equation}
W (\mathbf{r}, \mathbf{p}) = \int \frac{d^3 \mathbf{R}}{(2 \pi \hbar)^3} e^{-i \mathbf{p} \cdot\mathbf{R}/\hbar} \psi^*\left(\mathbf{r} - \frac{1}{2} \mathbf{R}\right) \psi\left(\mathbf{r} + \frac{1}{2} \mathbf{R}\right),
\label{nonrelwigner}
\end{equation}
which is a Fourier transform of a convolution of non-local wave functions. This phase-space distribution is extended to the relativistic scenario, where the wave functions in Eq.~(\ref{nonrelwigner}) are replaced by field operators that act on quantum states in Hilbert space. This concept was first studied in the QCD scenario by the authors Refs.~\cite{Belitsky:2003nz, Ji:2003ak}. The quark phase-space distribution is determined by the matrix element of the Wigner operator (at equal light-cone time $z^{+}$) 
\begin{eqnarray}
W^{q [\Gamma]}_{\lambda^{\prime} \lambda}(x,\mathbf{k}_{\perp}, \mathbf{b}_\perp, \mathscr{W}) &=& \int \frac{d^{2} \mathbf{\Delta}_\perp}{(2 \pi)^{2}} \left\langle p^{+}, \frac{\mathbf{\Delta}_\perp}{2}, \lambda^{\prime} \left|\hat{W}^{[\Gamma]}(x,\mathbf{k}_{\perp}, \mathbf{b}_\perp, \mathscr{W})\right| p^{+}, -\frac{\mathbf{\Delta}_\perp}{2}, \lambda \right\rangle,
\label{wignerfield}    
\end{eqnarray}
where the Wigner operator reads
\begin{equation}
\hat{W}^{[\Gamma]}(x,\mathbf{k}_{\perp}, \mathbf{b}_\perp, \mathscr{W}) = \frac{1}{2} \int \frac{d z^{-} d^{2} \mathbf{z}_\perp}{(2 \pi)^{3}} e^{i\left(x P^{+} z^{-} - \mathbf{k}_\perp\cdot \mathbf{z}_{\perp}\right)} \bar{\psi} \left(y - \frac{z}{2}\right) \Gamma \mathscr{W} \psi\left(y + \frac{z}{2}\right)\Bigg|_{z^{+} = 0},
\label{opwigner}
\end{equation}
$y^{\mu} = [0,0, \mathbf{b}_\perp]$, $\lambda (\lambda^{\prime})$ is the helicity of the incoming (outgoing) proton. The Wilson line is represented by $\mathscr{W}$ and ensures color gauge invariance of the interaction process. The vertex interaction information is specified by $\Gamma$ (Dirac operator): different choices of $\Gamma$ project out distinct quark spin and initial-state polarization structures, generating a full set of quark Wigner distributions. The off-forward proton states, with transverse momenta $\pm \mathbf{\Delta}_\perp/2$, encode the spatial resolution of the probe. The integration over the transverse momentum transfer $\mathbf{\Delta}_\perp$ introduces the dependence on the impact parameter: $\mathbf{b}_\perp$. The variables of the Wigner distribution are the longitudinal momentum fraction $x$, the intrinsic transverse momentum $\mathbf{k}_\perp$, and the transverse spatial position of the partons $\mathbf{b}_\perp$ with respect to the hadron's longitudinal center of mass. 

The Wigner distribution in Eq.~(\ref{wignerfield}) provides a five-dimensional tomography (two position and three momentum coordinates) of the hadron, as seen in the infinite-momentum frame \cite{Ji:2003ak, Belitsky:2003nz, Lorce:01}. The Wigner distribution depends on the average momentum and position of the partons. Hence, the Heisenberg uncertainty principle prevents the Wigner function from being a joint probability distribution, but allows this interpretation in the marginalized distributions of the Wigner distribution. The two known limits are the generalized parton distributions (GPD) and the transverse momentum distributions (TMD) \cite{introgpdtmd}. 
The GPDs are a limit of the Wigner distribution defined by integrating (or marginalizing) over $\mathbf{k}_\perp$, and thus yielding the distribution in impact parameter space $F(x,\mathbf{b}_\perp)$, which gives the probability density of finding a parton with a longitudinal momentum fraction $x$ at a transverse distance $\mathbf{b}_\perp$ from the center of momentum of the hadron. 
On the other hand, the TMDs $F(x,\mathbf{k}_\perp)$ are defined by marginalizing over the impact parameter in the Wigner distribution. Thus, the probability density of finding a parton with longitudinal momentum fraction $x$ and transverse momentum $\mathbf{k}_\perp$ is obtained. 
If one seeks to extract physically meaningful information from phase-space, the non-positive-definite nature of the Wigner distribution must be addressed. The best one can do is describe the system by probabilities of finding the particle inside the band $(q \pm \sigma_q/2, p \pm \sigma_p/2)$, with minimal uncertainty $\sigma_q \sigma_p = \hbar/2$. In this context, the Husimi distribution \cite{hattawherl,kuni1} arises as a natural candidate, since it is constructed via the Weierstrass transform, a Gaussian smearing of the Wigner function, thereby providing a positive-definite phase-space distribution. The Husimi distribution is defined as:
\begin{equation}
\label{wherldef}
H_{\ell}(x, \mathbf{r}, \mathbf{p})= \frac{1}{\pi^{2}}\int d^{2} \mathbf{r}^{\prime} d^{2} \mathbf{p}^{\prime} \exp\left[-\frac{(\mathbf{r}- \mathbf{r}^{\prime})^{2}}{\ell^{2}}\right] \exp \left[-\left(\mathbf{p} - \mathbf{p}^{\prime}\right)^{2} \ell^{2}\right] W(x,\mathbf{r}^{\prime}, \mathbf{p}^{\prime}),
\end{equation}
where $\ell$ is an arbitrary parameter with units of length. It is a parameter associated with the resolution of the system.
For the Husimi distribution parameterizing the nucleon, $\ell \lesssim R_h$, with $R_h$ being the hadronic radius. In the small-$x$ dominance regime, such as in the Color Glass Condensate model, $\ell = 1/ Q_s(x)$, where $Q_s(x)$ is the saturation scale \cite{Hatta:2015ggc}. As will later be discussed, $\ell$ will be related to the thermalization scale $\Lambda$ via $\ell \sim \Lambda ^{-1}$. In this case, $\langle \ln \left( H_\ell \right) \rangle$ is non-zero even for a pure state, because $H_h$ is a semiclassical distribution obtained from a coursed-grained description of the phase-space.
Having established a positive-definite phase-space distribution through the Husimi function, one can now consistently define entropy measures associated with the hadronic system. In particular, while the von Neumann entropy \cite{vonNeumann1955} vanishes for a pure state, such as the hadron wave function, the coarse-grained nature of the Husimi distribution allows for a non-trivial entropy. This motivates the introduction of the Wehrl entropy \cite{wehrl1979relation}, which provides a semiclassical measure of the information content in phase-space. In the following sections, we will explore how this is properly constructed and how it will be used to characterize the hydrodynamical initial state.

\section{Entropy formalism for High Energy Physics}
\label{sec:entropy_formalism_for_high_energy_physics}
\subsection{Wehrl Entropy}
\label{subsec:Shannon_Entropy}

Having established the relevant phase-space distributions, we now turn to the question of how to quantify the information they encode. A natural way to address this is through an entropic characterization, which provides a measure of the effective degrees of freedom contained in the system.

Before introducing the specific construction employed in this work, and its connection to Wehrl entropy, it is instructive to recall the classical notion of entropy in statistical mechanics. In this context, the Boltzmann-Gibbs entropy is defined as \cite{Gibbs1902}
\begin{equation}
S_{\text{BG}}=-k_B\int \frac{dpdq}{h^\prime}f(q,p)\ln f(q,p),
\label{bg-entropy}
\end{equation}
where $q$ and $p$ denote the generalized coordinate and momentum variables, and $h^\prime$ represents an elementary cell in phase space for the system described by the distribution function $f(q,p)$.
On the other hand we have  von Neumann entropy 
\begin{equation}
    S_{\rm vN} = - k_B \text{Tr}\left[\rho \ln \rho\right]
    \label{vonneuman}
\end{equation}
where $\rho$ is the density matrix operator of a given state $|\Psi \rangle$, that quantifies the entropy of a quantum system. However the connection of the quantum and classical definition is not trivial. Equation~(\ref{vonneuman}) does not reduce to the Eq.~(\ref{bg-entropy}) in the limit $h^{\prime}\rightarrow 0$. A. Wehrl thus proposed the notion of semiclassical entropy~\cite{wehrl1979relation}. Consider the coherent state $\ket{\lambda}$, with eigenvalue $\lambda=\frac{1}{\sqrt{2h}}(q+ip)$ of the annihilation operator, $a\ket{\lambda}=\lambda\ket{\lambda}$. Taking the trace of the {\it von Neumann} entropy, one obtains:
\begin{equation}
S_{\rm vN}=-\int \frac{dqdp}{2\pi\hbar}\bra{\lambda}\hat{\rho}\ln \hat{\rho}\ket{\lambda}.
\label{vn_coherent_state}
\end{equation}
The Wehrl entropy, $S_{\rm W}$, is obtained by performing the {\it classical substitution}, which consists of replacing $\bra{\lambda}\hat{\rho}\ln \hat{\rho}\ket{\lambda}$ with $\bra{\lambda}\hat{\rho}\ket{\lambda}\,\ln \bra{\lambda}\hat{\rho}\ket{\lambda}$. Thus,
\begin{equation}
S_{\rm W}=-\int \frac{dqdp}{2\pi\hbar}\bra{\lambda}\hat{\rho}\ket{\lambda}\,\ln\left(\bra{\lambda}\hat{\rho}\ket{\lambda}\right).
\label{classical_wehrl_entropy}
\end{equation}
Since $-x\ln x$ is a concave function, using Jensen’s inequality one finds $S_W>S_{vN}\geq 0$, and the equality $S_{\rm W}=S_{\rm vN}$ is impossible, $S_{\rm W}$ is always nonzero even for a pure state.

Now, considering a one-dimensional quantum system with a generic time-dependent pure state $\ket{\psi(t)}$, the Wigner distribution is defined as
\begin{equation}
W(q,p,t)=\int_{-\infty}^{\infty} dx\, e^{-ipx/\hbar}\,
\bra{\, q + x/2\,}\hat{\rho}(t)\ket{\,q - x/2\,},
\label{wigner_definition}
\end{equation}
where $\hat{\rho}(t)$ is the density matrix of a pure state. The Wigner distribution is a function of both position $q$ and momentum $p$, satisfying
\begin{equation}
\begin{cases}
\displaystyle \int dq\, \frac{1}{2\pi\hbar} W(q,p,t) = \left| \bra{\psi(t)} p \rangle \right|^2 ,\\[6pt]
\displaystyle \int dp\, \frac{1}{2\pi\hbar} W(q,p,t) = \left| \bra{\psi(t)} q \rangle \right|^2, \\[6pt]
\displaystyle \int dq\, dp\, \frac{1}{2\pi\hbar} W(q,p,t) = 1,
\end{cases}
\label{wigner_properties}
\end{equation}
with the last property ensuring proper normalization. The set of properties in Eq.~\eqref{wigner_properties} makes it tempting to interpret $W$ as a probability distribution in phase-space $(q,p)$. From Eq.~(\ref{wherldef}) we can define the Husimi distribution as a positive semidefinite quantity:
\begin{equation}
H(q,p,t)\equiv \bra{\lambda} \hat{\rho} \ket{\lambda} = \left|\langle \psi | \lambda \rangle\right|^{2} \ge 0.
\label{husimi_positive}
\end{equation}
This expression is essentially the trace of the density matrix in the $\lambda$ basis. As discussed in the previous section, the Bjorken variable $x$, the transverse momentum $\mathbf{k}_\perp$, and the impact parameter $\mathbf{b}_\perp$ are all present in the Wigner/Husimi distribution. Therefore, the Wehrl entropy is written as
\begin{equation}
S_{\rm W}(x) \equiv - \int d^{2}\mathbf{b}_{\perp}\, d^{2}\mathbf{k}_{\perp}\, xH(x,\mathbf{b}_{\perp},\mathbf{k}_{\perp})\, \ln[xH(x,\mathbf{b}_{\perp},\mathbf{k}_{\perp})],
\label{wehrl_qcd}
\end{equation}
we are defining the entropy per unit of rapidity, i.e., we use $xH$ \cite{hattawherl}. So far, we have defined how to construct the parton distribution inside hadrons, in particular its definition is terms of the phase-space distribution. This section then proceeded to devise a framework to quantify the entropy of the phase-space distribution of the hadrons. The missing ingredient is the connection of our formalism to relativistic fluid dynamics and its initial conditions, which we turn to next. 

\section{Relativistic Heavy Ion Collisions\label{initialhydro}}
The full characterization of a physical system with many degrees of freedom is, in general, a highly non-trivial task. However, an effective macroscopic description of this many-body system is possible by taking into account only the relevant degrees of freedom at a given scale. Such a macroscopic description is possible in the situation where the microscopic variables fluctuate rapidly in spacetime compared to macroscopic ones, hence the latter can be constructed by averaging over the former. Fluid dynamics is one example, where averaging over microscopic quantities leads to a system of equations which are governed by conserved macroscopic quantities that are then used to describe the system as a whole. In other words, fluid dynamics is an effective theory describing the long-wavelength, low frequency limit of the underlying microscopic dynamics of a system.

A fluid is defined as a continuous system in which every cell is assumed to be at local thermodynamic equilibrium. It is applicable when the fluid cell is large enough, compared to microscopic distance scales, to guarantee the proximity to thermodynamic equilibrium, and, at the same time, must be small enough, relative to the macroscopic distance scales, to ensure a continuum limit is achieved.
Relativistic hydrodynamics has been quite successful in explaining the various collective phenomena observed in astrophysics, cosmology and high-energy heavy-ion collisions, which is the target of this work. In this section we shall explore the construction of the ideal and dissipative relativistic hydrodynamics for heavy-ion collisions.

\subsection{Notations and Conventions}

For the purposes of this work, we are using the metric tensor of flat Minkowski space, within the ``mostly minus'' prescription, $g^{\mu \nu} = \mathrm{diag}(+,-,-,-)$, and solely consider natural units where $\hbar = k_B = c = 1$. Symmetrization of indices for a rank-2 tensor $A^{\mu \nu}$ is given as $A^{(\mu \nu)} \equiv \frac{1}{2}\left(A^{\mu \nu} + A^{\nu \mu}\right)$, while antisymmetrization is defined as $
A^{[\mu \nu]} \equiv \frac{1}{2}\left(A^{\mu \nu} - A^{\nu \mu}\right)$. The later-defined flow velocity, will be given by a time-like normalized four-vector $u^\mu$, i.e., $
u^{\mu} u_\mu = 1$. The projector onto the 3-space orthogonal to $u^\mu$ is denoted by
$\Delta^{\mu \nu} \equiv g^{\mu \nu} - u^\mu u^\nu$. The projection of a four-vector $A^\mu$ onto the 3-space orthogonal to $u^\mu$ is then $ A^{\langle \mu \rangle} \equiv \Delta^{\mu \nu} A_\nu$. The symmetric and traceless rank-2 projector onto the 3-space orthogonal to $u^\mu$ is $\Delta^{\mu \nu}_{\alpha \beta} \equiv
\Delta^\mu_{\,(\alpha}\Delta^\nu_{\,\beta)} -\frac{1}{3}\Delta^{\mu \nu}\Delta_{\alpha \beta}$. The projection of a rank-2 tensor \(A^{\mu \nu}\) onto the 3-space orthogonal to $u^\mu$ is then defined as $ A^{\langle \mu \nu \rangle} \equiv \Delta^{\mu \nu}_{\alpha \beta} A^{\alpha \beta}$.
\subsection{Conserved Quantities\label{conquan}}
The net particle number currents $N_q^{\mu}$ (where $q$ indexes conserved charges in the system)  and the energy momentum tensor $T^{\mu \nu}$ are the central quantities in the fluid dynamics description. They can be decomposed with respect to the time-like normalized four-vector $u^{\mu}$ as
\begin{equation}
    \begin{aligned}
        N_q^{\mu} &= n u^{\mu} + q^{\mu}, \\
        T^{\mu \nu} & = e u^{\mu} u^{\nu} - (p + \Pi)\Delta^{\mu \nu} +  h^{(\mu}u^{\nu)} + \pi^{\mu \nu}.
    \end{aligned}
    \label{nmutmunu}
\end{equation}
The particle-number densities $n \equiv N_{q,\mu} u^{\mu}$ in the local rest frame of the fluid,  $q^\mu \equiv N^{\langle \mu \rangle}$ is the particle diffusion currents with respect to the flow velocity $u^\mu$,  $ e \equiv T^{\mu\nu} u_\mu u_\nu$ is the energy density in the local rest frame of the fluid, $p$ is the thermodynamic pressure, $h^\mu \equiv T^{\langle \mu \rangle \nu} u_\nu$ is the energy-momentum diffusion current relative to $u^\mu$ (often called heat flow), and $\pi^{\mu\nu} \equiv T^{\langle \mu\nu \rangle}$ is the shear viscous pressure stress tensor, while $\Pi$ the bulk viscous pressure. We note that away from the ideal limit, flow has to be chosen according to some criterion. Common choices include the Landau definition, also known as the Landau frame, i.e., $h_\mu=0$, and the Eckart definition (frame), i.e., the choice where one of the conserved quantum numbers doesn't have a diffusion current such that $q_\mu=0$, but other choices are possible \cite{denicol}. In sum, in the Landau frame, energy density flow is being followed, while in the Eckart frame, a conserved quantum number flow is being followed. In all cases, $q^{\mu} u_{\mu} = 0$, $h^{\mu} u_{\mu} = 0$, $\pi^{\mu \nu} u_{\nu} = 0$ and $\pi^{\mu\nu}g_{\mu\nu} = 0$ by definition.

The conservation equations for the quantities in Eq.~(\ref{nmutmunu}) are
\begin{equation}
    \begin{aligned}
        \partial_\mu N^{\mu} & = 0,\\ 
        \partial_\mu T^{\mu \nu} & = 0.
    \end{aligned}
    \label{convequ}
\end{equation}
The conservation equation together with the Equation of State (EoS) forms a closed system of equations for ideal fluids. The next section will briefly discuss how the entropy behaves in such fluid. Finally we shall explore how to treat the dissipative hydrodynamics as well as some general features of entropy content. 
\subsection{Entropy Current in Relativistic Ideal Hydrodynamics \label{idealhydro}}

In the ideal fluid limit, i.e., in local thermodynamic equilibrium, we can define the EoS by expressing the thermodynamic pressure as a function of the local energy density and conserved quantum number density, $p = p(e, n)$, where $e$ and $n$ are obtained as coefficients from the irreducible tensor decomposition of $T^{\mu\nu}$ and $N^\mu$ shown in the previous section. In the ideal fluid limit, $n_\mu=\Pi=\Pi_{\mu \nu}=h_\mu=0$ and Eq.~(\ref{nmutmunu}) reduces to
\begin{equation}
    \begin{aligned}
    \label{tensorform}
        N_0^{\mu} &= n u^{\mu},\\
        T^{\mu \nu}_0 &= e u^{\mu}u^{\nu} - p \Delta^{\mu \nu}.
    \end{aligned}
\end{equation}
In this limit, the $e(x)$, $n(x)$, the fluid velocity $u^{\mu}(x)$ in Eq.~(\ref{convequ}), along with the EoS, completely define the system, thus forming a closed system of equations which are in principle solvable. For the purposes of this work we shall explore the thermodynamic entropy in the ideal fluid limit. In the covariant form, we can write the Euler and Gibbs-Duhem relations, respectively, as, 
\begin{equation}
    \begin{aligned}
    \label{gibbsduhem}
        S^{\mu}_0 &= p \beta^{\mu} + \beta_\nu T^{\mu \nu}_0 - \alpha N^{\mu}_0,\\
        d(p \beta) &= N^{\mu}_0 d \alpha - T^{\mu \nu}_0 \beta_\nu,
    \end{aligned}
\end{equation}
where $\alpha = \mu/T$, $\beta = 1/T$ and $\beta^{\mu} = \beta u^{\mu}$ for a system at temperature $T$. From these relations, the first law of thermodynamics becomes
\begin{equation}
    d S^{\mu}_0 = \beta_\nu d T^{\mu \nu}_0 - \alpha d N^{\mu}_0,
\end{equation}
which is used to write the divergence of the entropy current as
\begin{equation}
\label{entdiv}
    \partial_\mu S^{\mu}_0 = \beta_\mu \partial_\nu T^{\mu \nu}_0 - \alpha \partial_\mu N^{\mu}_0.
\end{equation}
Energy-momentum tensor conservation along with net particle number conservation, lead to entropy conservation $\partial_\mu S^{\mu}_0 = 0$ in the ideal limit. Thus, in equilibrium thermodynamics, entropy conservation is a natural consequence of the conservation of $T^{\mu \nu}_0$ and $N^{\mu}_0$. This is not surprising as thermal equilibrium is achieved whence entropy is maximized.
%

\subsection{Relativistic Dissipative Hydrodynamics and the Entropy Current\label{dissipativehydro}}

As previously discussed, the derivation of relativistic ideal fluid dynamics requires the local thermodynamic equilibrium, which is a strong condition. Most fluids in nature are non-ideal, i.e., the local thermodynamic equilibrium cannot be guaranteed and the deviations from this condition result in the dissipative effects. The latter are sources of entropy production, thus restoring thermal equilibrium for fluids that are close to such a state. 

The aim of this section is to briefly discuss the relativistic dissipative hydrodynamics and show how the dissipative effects are related to the divergence of the entropy current. In this limit, the full \eqn{nmutmunu} must be used, and expressions for the dissipative currents must $\mathcal{Q}=\Pi,n_\mu,\Pi_{\mu \nu}$ must be found. For a deterministic theory, one expects the ``first order correction'' to be of the form of the gradient of equilibrium quantities multiplied by a microscopic dissipative quantity $\ell_{\text{mfp}}$ (corresponding to the mean free path/sound wave dissipation length), $\mathcal{Q}=\order{\ell_{\text{mfp}}\, \partial (u_\mu,e,n)}$. However the diffusive nature of such dynamics makes the theory inherently acausal, necessitating a ``second-order'' Maxwell-Cattaneo type regularization of the form \cite{denicol}
\begin{equation}
\tau u_\mu \partial^\mu \mathcal{Q}+\mathcal{Q}=\ell_{\text{mfp}} \order{\partial (u_\mu,e,n)}+\order{\ell_{\text{mfp}}\times \mathcal{Q}^{2}},
\end{equation}
with the coefficients calculated from Kubo-type formulae and adjusted to ensure the positivity of the entropy current.

While most implementations of hydrodynamics involve solving the conservation laws, hydrodynamics can in principle also be formulated in terms of an entropy current.  
For dissipative dynamics, the second equation is invariant, while the first one becomes takes the form
\begin{equation}
\label{entropygen}
\partial_\mu S^\mu=\Pi \partial_\mu \left(\frac{u^\mu}{T}\right)+\pi_{\mu \nu}\partial^\mu \left( \frac{u^\nu}{T} \right)- n^\mu \partial_\mu \left(\frac{\mu_q}{T} \right),
\end{equation}
where $\mu_q$ is the chemical potential, which, for the sake of the discussion here, is associated with conserved net baryon number solely. Provided dynamics are relaxational, the entropy current is always positive and the second law of hydrodynamics is satisfied.

Because conserved currents are usually obtained via constitutive relations combined with relaxation-type dynamics, the Gibbs-Duhem relation is defined under conditions of entropy maximization, and any flow away from the ideal hydrodynamic limit is not uniquely defined. In other words, there is no general procedure to obtain an order-by-order expression of \eqn{entropygen} which matches the entropy current to the same order as the conserved currents. Therefore, what is usually done is to solve the conservation equations (which are of course exact and lead to well-behaved equations of motion) and adjust the entropy current via the Gibbs stability criterion \cite{gavastab}. For large systems whose closeness to local equilibrium is not in discussion, such a procedure is generally well founded \cite{geroch}. For small fluctuating systems which are also strongly coupled, this recipe is unreliable, precisely because the entropy encodes closeness to equilibrium, thus determining the distribution of fluctuations. In the next section, we shall review \cite{gencov} how to write a fluctuating dissipative hydrodynamics based on an entropy current, and explicitly link the entropy defined previously to one which can be used in this context.
\subsection{Relativistic Fluctuating Hydrodynamics \label{secfluct}}

In the previous section we defined the entropy content of the hadron through the coarse-grained phase-space distribution, and so it is necessary to establish how this information is mapped onto the framework of relativistic hydrodynamics. Thus we have to first discuss the issue of the quantum state (nucleon wave function) and its classical theory (hydrodynamics). Quantum mechanics is inherently contextual  \cite{bell}: the state of a physical system is described by an operator-valued object, such as a density matrix $\hat{\rho}$ or a Wigner functional. Experimental observables are not the state itself, but the expectation value $\langle \hat{O} \rangle = \text{Tr}[\hat{O} \times \hat{\rho}]$. This implies that non-commuting observables encode incompatible information, a challenge exemplified by the relationship between generalized parton distributions and transverse momentum distributions. The experimental observation of hydrodynamic-like behavior in small systems \cite{nagle,small} represents a challenge, since such systems involve only a few constituents and evolve over very short time scales. Hydrodynamics near the ideal limit is based on statistical mechanics \cite{zubbecc,crooks,ergo,gencov}, where macroscopic behavior emerges from ensemble averages over microscopic fluctuations. In this framework, the initial conditions of highly fluctuating hydrodynamic evolution can be consistently defined from an ensemble generated by a partition function $\mathcal{Z}$,
\begin{equation}
\label{zclass}
\mathcal{Z} \equiv \mathrm{Tr}\left[ \exp \left[- \int_{\Sigma} d\Sigma_\mu \left(\beta_\nu \hat{T}^{\mu \nu}-\mu \hat{N}^\mu\right) \right] \right]\eqcomma \left. d\Sigma_\mu \right|_{\mathrm{rest}}=\left( \begin{array}{c}d^{3} x\\ \mathbf{0} \end{array}\right),
\end{equation}
where $\beta_\mu$, $\beta$ and $u_\mu$ are defined under Eq.~(\ref{gibbsduhem}). The authors in Ref.~\cite{gencov} point out that assuming a Gaussian ansatz throughout the hydrodynamic stage
\begin{equation}
\mathcal{Z} \sim \exp\left[- C_{\mu \nu \alpha \beta}(\Sigma_{\gamma},\Sigma'_\zeta) \left(\hat{T}^{\mu \nu}(\Sigma_\gamma)- \ave{T^{\mu \nu} (\Sigma_\gamma ) }\right) \left(\hat{T}^{\alpha \beta}(\Sigma'_\zeta)- \ave{T^{\mu \nu} (\Sigma'_\zeta )} \right)  \right],
\label{gaussian}
\end{equation}
together with the hypothesis of general covariance under refoliations of $\Sigma^{\mu}$, is enough to specify the underlying system's dynamics.~\footnote{$C_{\mu\nu\alpha\beta}$ will be discussed shortly.} 
This, however, is a classical probability density, respectful of Bell's and Mermin's relations \cite{bell}, unlike $\langle \hat O\rangle$. Finding the right $O$ to match a classical operator of the type in \eqn{zclass}, under the assumption of \textit{fast thermalization} (motivated by the eigenstate thermalization hypothesis~\cite{eth}) is then the problem to be addressed in this work.

Although the $\Lambda$ scale, and the constant of proportionality, are not calculable from first principles, a link between the Wigner function to the Gaussian partition function defined in Ref.~\cite{gencov} can be established. The correlator of the energy-momentum tensor of Ref.~\cite{gencov}, defines the (volume)$\times$(time) scale where it makes sense to talk about a maximally mixed phase space distribution, $C\equiv C_{\alpha \beta}^{\alpha \beta} \sim \Lambda^{-8}$. Thus, \eqn{gaussian} would constrain the initial condition for \cite{gencov} together with Eq.~(\ref{hydroI}) as 
 \begin{equation}
 C_{\alpha \beta \mu \nu} (\tau_0) = N \frac{\ave{d^{2}\ln \mathcal{H}}_{\Lambda}}{d\Lambda^{-2}} \left( g_{\alpha \mu }g_{ \beta \nu}+g_{\alpha \nu} g_{ \beta \mu}-g_{\alpha \beta} g_{\mu \nu } \right),
 \end{equation}
where the initial flow $u_\mu$ and co-moving geometry $g_{\mu \nu}$ are given by the Milne-like coordinates with 1 dimensional Bjorken flow (with the initialization time $\tau_0\equiv \Lambda$ the coarse-graining scale)\cite{martinez}
 \begin{equation}
 u_\mu =\left(  \begin{array}{c}\cosh(y)\\ 0 \\ 0 \\ \sinh(y)
 \end{array}\right) \eqcomma y=\pm \tanh^{-1}\left( x^{-1} \right).
 \end{equation}
 and  the coorindate system is defined by
 \[ x_\mu=\left(\tau=\sqrt{t^{2}-z^{2}},r,\phi,y\right)  \eqcomma g_{\mu \nu}= \left( \begin{array}{cccc}
 1 & 0 & 0 & 0\\
  0   & -1  & 0 &0\\
  0   & 0 & -r^{2} & 0\\
  0 & 0 &0  & -\tau^{2}
 \end{array} \right)\]
An estimate of the entropy carried by such a partition function can therefore be identified with the Wehrl entropy partition thermalized on a spacetime scale given by the Husimi smearing parameter, $\ave{\ln \mathcal{H}}_\Lambda$ (the definition of $\Lambda$ in a field theory is discussed in the next section). By computing this entropy, we provide a self-consistent bridge between the initial quantum state and the classical hydrodynamics (following for instance Eq.~(\ref{targetproj})), which in the context of \cite{gencov} would properly characterize the collectivity in small collision systems.

 The approaches of this subsection and the previous one underpin the difference between the attractor picture \cite{attractor,attractor2} and the approach based on the fundamental applicability of statistical mechanics, inspired by the entanglement entropy picture \cite{kharzeev,popescu} and Eigenstate thermalization \cite{eth}.  In the former case, one demands that the elements of the energy momentum tensor look close to the average values of a thermalized ensemble, but no assumption is made regarding the distribution of microstates and how close is the system to ``real'' thermalization.   In the latter, hydrodynamization and thermalization coincide in the sense that all microstates are assumed to be equally likely and unentangled (as in Eigenstate thermalization), but, it does not mean that the system is always in it's expectation value, just that the probability for the density to be in a certain value follows a statistical distribution of microstates.   If thermalization is ``fast'', the symmetries of the light-cone frame make it inevitable that partonic entanglement entropy \cite{kharzeev}, characterizing the maximally mixed state of the hadron \cite{popescu}, dominates its total entropy, since additional entropy production would come associated with a time-scale resulting in a breakdown of collectivity in systems whose size is smaller than that scale.
 
 It is clear that only the latter scenario can give a consistent picture of hydrodynamic fluctuations, which in fact tend to spoil the highly symmetric structures in which the existence of attractors is usually demonstrated.   While there is experimental evidence that the second picture provides a good description of fluctuations in experimental data \cite{kharzeev2,kutakcaputa,kutaklokos}, there is no conclusive evidence that this extends to transverse space, as required by the scenario developed in this subsection.  In the next section, we will provide the recipe for developing such a quantitative comparison, by linking partonic structure to both hydrodynamic entropy density and observable quantities. 
\section{Initial Entropy Deposition Model\label{modelmain}}
We now have all the ingredients needed to connect the coarse-grained Wherl entropy, the quantum objects encoding the microscopic partonic structure and the classical object encoding information for hydrodynamics.

The latter, defined using the Fluctuating Ansatz described in the last section, is a Gaussian statistical distribution.   The Husimi procedure can be used to transform a quantum mechanical object into exactly such a distribution.  Missing, however, are two ingredients: How to generalize Husimi entropy to systems where the number of degrees of freedom is not set, and how to connect the entropy to the partonic structure.  The next two sections will address these two issues in detail.
\subsection{From partonic structure to entropy}
In this section, we explore the characterization of the initial entropy deposition based on the phase-space nucleon entropy, in particular a Wehrl-\textit{like} entropy. As discussed in the previous section, hydrodynamics requires the system to be in, or close to, local thermal equilibrium, which in quantum mechanical terms corresponds to a maximally mixed state. In this work, a maximally mixed state should be understood in a coarse–-grained phase–-space sense, where microscopic quantum coherences are averaged over a resolution scale (this discussion will be clarified further in the next section). 

In Section \ref{GPD}, we examined generalized parton distributions (GPDs), transverse momentum distributions (TMDs), Wigner and Husimi distributions. Among these, the phase-space distribution provides the most suitable description once coarse-–grained, as it can be interpreted as an effective mixed state. Motivated by this observation, we first discuss the limitations of TMDs and GPDs in describing the initial energy and entropy deposition. We then formulate an appropriate model for the initial conditions based on the phase-space nucleon entropy.

In the paradigm of hydrodynamization \cite{attractor,attractor2}, the onset of hydrodynamic behavior does not necessarily imply local thermalization (nor isotropization) of the system. In this regime, the initial conditions may be characterized by strongly out-of-equilibrium momentum distributions, such that the relevant macroscopic description is formulated in terms of the energy-momentum tensor. In other words, hydrodynamic behavior can emerge even when the system is far from a thermal (or isotropized) state. In contrast, if the system were locally thermalized, it would correspond to an effectively maximally mixed state in a coarse-grained sense. In this case, the appropriate macroscopic quantity is the entropy current rather than the energy-momentum tensor. The entropy then provides a more natural measure of the available phase-space for the degrees of freedom, which is precisely the quantity we aim to characterize in this work.

In the first scenario, TMDs would provide a natural microscopic description of the initial energy deposition, from which the energy-momentum tensor can be constructed and evolved following Eq.~(\ref{convequ}). However, the expectation value of the TMD-informed energy-momentum tensor may violate energy conditions \cite{jorgeec:01}. On the other hand, in the second scenario, since entropy production in ultrarelativistic systems is closely related to the local density of degrees of freedom, one is naturally led to consider GPDs as candidates for the initial entropy deposition. However, GPDs describe the ground-state structure of a single hadron and do not correspond to an ensemble. Therefore, GPDs cannot be directly interpreted as an entropy density. Therefore, considering the system as already thermalized, the relevant initial state should behave as a maximally decohered state \cite{popescu,kharzeev,kharzeev2}. Experimentally, this has been verified \cite{kharzeev3} for point-like deep inelastic scattering events \cite{kutakcaputa,kutaklokos} and $pp$ collisions, where the entropy inferred from the parton wave function of the nucleon \cite{brodsky} (via \eqn{psidef} and Eq.~(3.17) of \cite{brodsky}, \eqn{gdef}) matches the measured value $\ave{\ln P(N)}$, with $P(N)$ the distribution of multiplicities. Such an agreement can be reasonably assumed to work for point-like process, where the scattering probes a localized portion of the wave function. However, when the interacting system has a finite spatial extent, hotspots (i.e. spatial fluctuations) spoil this direct correspondence between the wave functions and the particle multiplicity. This scenario is where one expects the effective thermodynamic description to transition towards hydrodynamics. 

The central assumption of this work is that the coarse-grained phase-space distribution of the nucleon provides a valid proxy for the initial entropy density relevant for hydrodynamic evolution. In this picture, the loss of phase information associated with coarse-graining  leads to an effectively mixed description of the system in phase space. As a result, the appropriate object is a Husimi-\textit{like} distribution, which is positive definite and admits a natural entropic interpretation. The corresponding entropy can therefore be consistently defined through a Wehrl-\textit{like} entropy, which quantifies the number of accessible phase-space degrees of freedom at a given resolution scale.

To explicitly construct such a coarse-grained phase-space distribution, we start from the parton distributions in the Light-Cone Wave Function (LCWF) formalism \cite{brodsky}, and generalize them by introducing an explicit transverse spatial dependence. This is achieved by performing a Fourier transform of the transverse coordinates of the parton distribution function (Eq.~(3.17) of \cite{brodsky}). Using Parseval's theorem, the gluon structure function with scale dependence $Q^2$ can be re-expressed as
\begin{equation}
\label{gdef}
G(x,Q^2)= \sum_{n,\lambda_i} \int \prod dx_i d^{2} \mathbf{r}_{\perp i}\left| \tilde{\Psi}^Q\left(\mathbf{r}_{\perp i},x_i,\lambda_i \right)\right|^{2}\sum_i\delta\left(x-x_i\right),
\end{equation}
where $\tilde{\Psi}^Q$ is the Fourier transform of $\Psi^Q\left(\mathbf{k}_{\perp i},x_i,\lambda_i\right)$ in the LCWF. Counting the partons in a transverse cell $d^{2} \mathbf{b}_\perp $ one obtains
\begin{equation}
\label{rhodef}
\bar{G}(x,\mathbf{b}_\perp,Q^{2})= \sum_{n,\lambda_i} \int \prod dx_i d^{2} \mathbf{r}_{\perp i}\left| \tilde{\Psi}^Q\left(\mathbf{r}_{\perp i},x_i,\lambda_i \right)\right|^{2}\sum_i\delta\left(x-x_i\right)\delta^{2}\left( \mathbf{r}_\perp - \mathbf{b}_{\perp i}\right).
\end{equation}
Like \eqn{gdef}, \eqn{rhodef} is independent of the phase of the gluon fields, and hence according to \cite{kharzeev} it qualifies as a maximally decohered quantity which can serve as a basis for an initial entropy density. The $Q^{2}$ dependence of \eqn{rhodef}, just like in \eqn{gdef}, comes from renormalization group running of LCWF, and is therefore $\sim \ln Q^{2}$.  Something similar was calculated in \cite{mullerfilm} from the point of view of a hadron at rest being trasversed by a highly boosted ``pancake``.

It is worth comparing this object to the usual description of a transverse parton distribution of the nucleon, the GPD.
In fact, GPDs can be written using light cone wave function (LCWF), as early studies show \cite{Diehl:2000xz}.  At finite $\xi$  GPDs can even be written as an overlap of wave functions, 
\begin{equation}
    \mathcal{H}^{q (N \rightarrow N)}_{\lambda^{\prime} \lambda} =  \lim_{\xi \rightarrow 0} \sqrt{1 - \xi}^{1-N} \sqrt{1 + \xi}^{1-N} \sum_{\beta = \beta^{\prime}} \sum_j \delta_{s_j q} \int [d x]_N [d^{2} \mathbf{k}_\perp]_N  \delta(x - x_j) \Psi^{* \lambda^{\prime}}_{N, \beta^{\prime}} (\hat{r}^{\prime}(\xi))\Psi^{\lambda}_{N,\beta}(\tilde{r}(\xi)),
    \label{overlaprepresentation}
\end{equation}
where N is the number of parton Fock states, $\tilde{r}(\hat{r}^{\prime})$ for the incoming (outgoing) proton and $\xi$ tracks the longitudinal momentum exchange (the nucleon is on-shell for $\xi\rightarrow 0$).  At the experimentally inaccessible $\xi \rightarrow 0$, this limit is positive-definite, making the structure of \ref{overlaprepresentation} analogous to \eqn{rhodef}.  It is however contradictory to interpret such a limit as an entropy, because it corresponds to a pure state which has not experienced any interaction.   A finite $\xi$ state, on the other hand, is mixed, and not necessarily positive definite, so $Q\times \xi$ can not be used as a coarse-graining scale to match the initial partonic structure to a hydrodynamic entropy.  

To summarize this section, we have shown how, under conditions of ``fast'' Eigenstate thermalization, the entanglement entropy dominance implicit in \cite{kharzeev} can be used to obtain not just a total entropy but a transverse entropy density usable for fluctuating hydrodynamics.   Of course, impact-parameter dependent structure functions are not experimentally measureable, thus the same methods of \cite{kharzeev} can not be used to predict hydrodynamic initial conditions. In fact, the comparison of \eqn{rhodef} to GPDs shows that to link a ``fast thermalization hypothesis'' a la \cite{kharzeev} to experimental data we need an object that can be matched to \eqn{rhodef} but with a coarse-graining scale that preserves the positive-definiteness of \eqn{overlaprepresentation}. As we show in the next section, Husimi type coarse-graining can be used to relate this entropy density to an object of the same form of that obtained in Section \ref{secfluct}, with the scale adjustable to global multiplicity.   
\subsection{The Wigner function and Entropy Content of the Hadron} \label{wignerentropycontent}
In this subsection, we provide a connection between the coarse-grained phase-space description of the nucleon and the entropy relevant for hydrodynamic initial conditions. In field theory, the canonical degrees of freedom are the infinite Fourier coefficients of the field, whose detection is in addition limited by the scale accessible to the experiments. Thus, the closest definition to \eqn{nonrelwigner} is $\mathcal{W}$, the Wigner {\em functional} \cite{wigfunc}, in terms of a generic field variable $\phi$ ($\phi\equiv A^\mu_i$ for QCD) and its canonical momentum $\Pi$, is
\begin{equation}
\label{wigfuncdef}
\mathcal{W}_\Lambda(\phi,\Pi)=\int \mathcal{D}\phi' \exp \left[ i\int^\Lambda dp \phi'(p) \Pi(p) \right]    \bra{\phi+\frac{1}{2}\phi'}\hat{\rho} \ket{\phi-\frac{1}{2}\phi'} ,
\end{equation}
where a momentum cutoff $\Lambda$ is added, thus incorporating the scale to which experiments are sensitive.\footnote{Note that the integral $\mathcal{D}\phi$ is done only up to $\phi$s of momentum $\leq \Lambda$.} This is the fundamental difference between quantum mechanics and quantum field theory, and it has profound consequences on how entropy is defined \cite{nishioka}.   Computationally, the limit $\Lambda \rightarrow \infty$ leads to a divergence of the Husimi/Wehrl entropy $\lim_{\Lambda \rightarrow \infty} \ave{\ln \mathcal{W}}_{\Pi} \sim \Lambda^{3}$, reflecting the ultraviolet proliferation of degrees of freedom in QFT. This divergence does not imply that the coarse-grained Husimi entropy can be regularized into the von Neumann entropy \cite{vonNeumann1955} of the pure quantum state. Rather, it indicates that the Husimi distribution represents a semiclassical distribution whose entropy increases with resolution. After renormalization, it is possible to define an entropy density, but that entropy density is not the fundamental von Neumann entropy. This is an illustration of the delocalization and intercorrelation of quantum field theory states, while also providing a relation between entropy and quantum mechanics \cite{carcassi}.

An important issue that needs to be resolved is the role of $\ell$ (as defined in \eqn{wherldef}) in the Wigner functional picture. Unlike in \eqn{wherldef}, \eqn{wigfuncdef} does not have a well-defined number of degrees of freedom, and  $(\phi,\Pi)$ acquire a dimension of energy squared and its inverse, respectively, for bosonic fields.
According to the paradigms of the renormalization group/effective theory, one should be able to obtain a consistent limit by regularizing the theory with a parameter $\Lambda$ (in the integrand of the action of \eqn{wigfuncdef}) and expect a smooth flow of any observable, or lagrangian parameter, with $\Lambda$ which depends only on $\Lambda$ and ``bare'' parameters that can be fixed to observables of an experimental scale (for relevant and marginal operators there is a well-defined fixed point for $\Lambda \rightarrow \infty$, for irrelevant ones this limit is divergent).
One physically straight-forward way to see this regularization is via a lattice, with link size $\Delta x \sim \Lambda^{-1}$ so the problem is reduced to a finite quantum mechanics problem
\[\ \Delta x \sim \Lambda^{-1} \eqcomma \mathcal{D}\phi \mathcal{D}\Pi \rightarrow \int \prod_{k=0}^\Lambda d\phi_k d\Pi_k    \]
Now, choosing $\ell \ne \Lambda^{-1}$ will mean there will be two dimensionful parameters regulating Husimi coarse-grained dynamics, which will, in general, make any observable sensitive to $\ell \Lambda$, a dimensionless parameter with no physical significance.
This makes it clear that identifying $\ell$ of \eqn{wherldef} with $\Lambda^{-1}$ keeps the Wherl entropy consistent as the lattice scale varies, including the continuum approach, while keeping the entanglement entropy scaling with the form expected from quantum field theory \cite{nishioka} (where ``total entropy'' is indeed an irrelevant operator, through an unobservable one at low energies, even for super-renormalizeable theories). This $\Lambda$ is, however, a very different object from that of the renormalization group, since the Husimi coarse-graining produces entropy and heat with respect to the pure quantum state. Decreasing the lattice size $x\rightarrow x'$ means increasing the entropy density everywhere by the appropriate dimensional factor $(x/x')^3$ while maintaining the distribution intact. Physically, what we are doing is assuming that in the initial strongly coupled regime the field undergoes Eigenstate thermalization \cite{eth} over the time-scale of a lattice size, so while a continuum evolution to the continuum is possible, the continuum will have an infinite entropy, consistent with the fact that a continuum QFT has an infinite number of degrees of freedom.
$\Lambda$ is therefore a physical scale which will appear in observables, and, as we shall see, needs to be fitted to the overall multiplicity.

For an extended system where local equilibrium and hydrodynamics are expected to be relevant, $\Lambda$ is determined by the energy budget available in the local thermalized cell. The initial state of this cell is, in turn, bounded below by its size via the uncertainty principle. Therefore a suitable generalization of \eqn{wherldef} is
\begin{equation}
\label{wherlqft}
\mathcal{H}_{\Lambda}(\phi,\Pi)= \int  \mathcal{D}\phi' \mathcal{D}\Pi' \exp\left[-\Lambda^{2} (\phi-\phi')^{2}\right] \exp \left[-\frac{(\Pi-\Pi')^{2}}{\Lambda^{2}}\right] \mathcal{W}(\phi',\Pi'),
\end{equation}
the Husimi functional, where the localization scale $\Lambda$ hence enters into the Gaussian smearing.
An entropy density $\sim \Lambda^{3}$, as required \cite{nishioka}; $\ave{\ln \mathcal{H}}_\Lambda$ is an estimate for the entropy of a field whose degrees of freedom are localized in cells of size $\Lambda^{-1}$.
Connecting \eqn{wherlqft} to the result in Eq.~(\ref{rhodef}), and motivated by the entanglement entropy in \cite{kharzeev,kharzeev2,kharzeev3}, we can use the parton multiplicity as a proxy for the entropy density. Thus, we can define a Wehrl-like entropy of the quantity \eqn{rhodef} as
\begin{equation}
\ave{\ln \mathcal{H}}_{\Lambda}\left(x,\mathbf{b}\right) = \mathcal{N} \int_0^\Lambda d^{2} Q^{2}  \, \bar{G}(x,\mathbf{b}_\perp,Q^{2}),
\label{qq}
\end{equation}
The physical interpretation of this equation is that the localization of degrees of freedom, described by the probability distributions in \eqn{rhodef}, is a good proxy for the Husimi smearing of the Wigner functional of the coherent hadronic state. Probing parton degrees of freedom at $Q^{2}$ breaks the coherence of the Wigner functional at that scale, producing  an amount of entropy approximated by the Husimi distribution. Importantly, the later entropy is not the von Neumann entropy of the nucleon's ground state. The entropy in Eq.~(\ref{qq}) does not quantify fundamental quantum entanglement; rather, it reflects the information loss associated with the coarse-grained Husimi distribution.

Assuming that this coarse-graining-driven decoherence brings the system close to equilibrium, $\Lambda$ can be set from the initial values of the energy-momentum tensor
 \begin{equation}
\ave{T}_{\mu \nu}(\tau_0)=eu_\mu u_\nu - p \Delta_{\mu \nu} \eqcomma e= \frac{\ave{d\ln \mathcal{H}}_{\Lambda}}{d\Lambda} \eqcomma p=p(e).
\label{hydroI}
 \end{equation}
Given a model for the Wigner function of the nucleon, the total multiplicity of a \textit{pp} or a \textit{pA} collision is sufficient to set the scale $\Lambda$ of the entropy. Since, within our ansatz, the produced entanglement entropy is extensive, the target and projective entropies add up. Using the Bjorken picture, this means that the the projectile $P$ and target $T$ are combined as
\begin{equation}
\label{targetproj}
\ave{\ln \mathcal{H}}^{\text{IC}}_{\Lambda}\left(x,\mathbf{b}_\perp\right)  = N \int dx_{P,T} \left[ \ave{\ln \mathcal{H}}_{\Lambda}^{P}\left(x_{P},\mathbf{b}_\perp\right) +  \ave{\ln \mathcal{H}}_{\Lambda}^{T}\left(x_{T},\mathbf{b}_\perp\right) \right] \delta \left( x_{P}+ x_{T} -x \right).
\end{equation}
The quantity $\ave{\ln \mathcal{H}}^{\text{IC}}_{\Lambda}\left(x,\mathbf{b}_\perp\right)$ therefore defines the initial entropy deposition, based on the phase-space nucleon entropy. 
Using the relation between entropy and final multiplicity \cite{Florkowski:2010zz} one can therefore estimate that
\begin{equation}
\label{lambdafix}
    \mathcal{H}^{\text{IC}}_{\Lambda} \sim 4 \Lambda^3 \Delta y \frac{dN}{dy}.
\end{equation}
Note that the summation of the two terms follows from extensivity, which is a consequence of the dominance of entropy by entanglement entropy inherent in Eigenstate thermalization.  Implicit in the above discussion is the assumption that $\Lambda$, fitted from multiplicity, is parametrically larger than $1/R_h$, the radius relevant from nuclear structure. If not, fluctuations will smear out any flow observable. As we shall now see this has qualitative experimental consequences.

The result in Eq.~(\ref{targetproj}) can be used as an estimate of the eccentricity $\epsilon_n$ and then used to probe the anisotropic flow $v_n$ in $pp$ and $pA$ collisions as calculated from an initial Wigner function, including {\em polarized} $p^\uparrow p$ and $p^\uparrow A$ collisions.   By a simple dimensional analysis \cite{lacey}, expanding in small dimensionless quantities one expects
\begin{equation}
\label{v2eps2}
\frac{v_n}{\epsilon_n} \sim A\times \left(1- B(R,T) n^{2} \mathrm{Kn}\right)+\sum_m C_m(R,T) \epsilon_m+ \order{\epsilon \times \mathrm{Kn,\epsilon^2}},
\end{equation}
where $\mathrm{Kn}$ is the Knudsen number, $R,T$ are the initial temperature and characteristic size (parametrizing the lifetime) while $A,B,C_m$ are dimensionless coefficients, for which one needs a full hydrodynamic simulation to compute. Given the basic scaling of eccentricity of Glauber type dynamics ($\sim n$; note the difference with the dissipative scaling $\sim n^{2}$ \cite{lacey}), we expect fluctuations to overwhelm initial state information when $\Lambda \sim n/R_h$ where $R_h$ is the nucleon size.   This can be used to derive a lower bound for multiplicity in the applicability of our model
\begin{equation} \label{multmin} v_n\left(\frac{dN}{dy} \leq \left. \frac{dN}{dy}\right|_{min} \right) \ll \left. v_n\right|_{hydro} \eqcomma  \left. \frac{dN}{dy}\right|_{min} \sim n \times \frac{1}{R_h \times \Lambda}\end{equation}
testable if the sort of correlations seen in low multiplicity events \cite{alice} would be translated into Fourier components, or in cumulant analysis of low multiplicity events \cite{cms1,cms2,atlas}.  A quantitative test, of course, needs a numerical estimate, which is beyond the scope of this work.  However, it is clear that once $\Lambda$ is fixed via \eqn{lambdafix} \eqn{v2eps2} and \eqn{multmin} can be used to falsify the fast thermalization entanglement entropy dominated picture described in this work.
Another generic qualitative prediction is that $p^\uparrow A$ collisions and $pA$ collisions should have somewhat different $v_n$ in bins of equal multiplicity. To make this quantitative, one needs an estimate of the {\em Wigner function}, and in particular its quark contribution (large-$x$ regime), as done in \cite{Lorce:01} for instance. Formulating the equivalent equation to \eqn{targetproj} using GPD parton multiplicities in \textit{lieu} of $\bar{G}(x,\mathbf{b}_\perp,Q^{2})$ leads to incorrect physics that neglects the intrinsic contribution to entropy from disentanglement, as explored in \cite{kharzeev}.

\section{Discussion and Conclusions \label{discussion}}

The proper formulation of relativistic hydrodynamics to small collision systems like $pp$ or $pA$ requires a consistent matching between the microscopic quantum description of hadrons and the macroscopic, effectively classical, framework of fluid dynamics. This matching is nontrivial, since the nucleon is described by a pure quantum state, while hydrodynamics assumes an entropy current associated with a maximally mixed ensemble.
In this work, we have proposed that this mismatch is naturally resolved through a coarse-graining of the nucleon phase-space entropy. Starting from the Wigner distribution, we showed that the appropriate object for describing hydrodynamic initial conditions is a Husimi-like distribution for the nucleon, which incorporates a finite resolution scale and yields a positive-definite phase-space density. The associated Wehrl entropy of the coarse-grained nucleon distribution then provides a semiclassical measure of the number of accessible degrees of freedom, and therefore a natural candidate for the initial entropy density required by hydrodynamics.

Within this framework, we demonstrated that commonly used partonic observables such as GPDs and TMDs are insufficient to characterize entropy deposition, as they correspond to projections of a pure state and lack the necessary coarse-grained structure. Instead, the entropy must be constructed from phase-space distributions that explicitly encode both spatial and momentum information at a finite resolution scale. This leads to a consistent interpretation of entropy production as arising from the loss of phase information under coarse-graining, rather than from fundamental quantum entanglement. However, in principle, the lattice can provide a direct link to the Wigner function in the future \cite{lattice3d}.  If this is the case, then given global multiplicity the model described here, together with a code implementing the effective theory of \cite{gencov} could directly link lattice data with correlations traditionally associated with hydrodynamics.

A key outcome of this approach is that the initial entropy density can be related to parton and hadron multiplicities integrated up to a scale set by the localization parameter $\Lambda$, which plays the role of a thermalization scale. This establishes a direct link between the microscopic structure of the nucleon and the macroscopic quantities required to initialize hydrodynamics, providing a self-consistent bridge between quantum field theory and fluid dynamics in small collision systems.
The framework proposed here is testable. In particular, it predicts that observables sensitive to the spatial structure of the nucleon, such as anisotropic flow coefficients in $pp$ and $pA$ collisions, should reflect the underlying phase-space distribution rather than its projections. Extensions of this approach to polarized collisions offer a promising avenue to probe the role of spin-dependent spatial anisotropies in the initial state, which would be a interesting probe of the discrepancy of the anisotropy flow in $pA$ collision as showed in \cite{phenixv2,phenixv3}.

Finally, this work highlights the importance of resolution scale and coarse–graining in defining entropy in quantum field theory. The Husimi smearing parameter $\Lambda$ encapsulates the transition from a coherent quantum description to an effectively classical ensemble, suggesting that hydrodynamization can be understood as a process of phase-space decoherence governed by the experimental and dynamical resolution scales. Further quantitative implementations of this framework, including realistic models of the Wigner distribution, will allow for direct comparisons with experimental data and provide deeper insight into the emergence of collectivity in small systems.

\section*{ACKNOWLEDGMENTS}
G.R.S. acknowledge the hospitality at University of Regina where part of this work was conducted. G.R.S. acknowledges the Emerging Leaders in the Americas Program scholarship, funded by Global Affairs Canada, for the financial support provided during the visit to UofR. G.R.S. is supported by the CAPES doctoral fellowship 88887.005836/2024-00. G.T. thanks Bolsa de produtividade CNPQ 305731/2023-8 and FAPESP 2023/06278-2, as well as FAPESP temático 2023/13749-1 for support. G.V. is grateful for the support provided by the Canada Research Chair under grant number CRC-2022-00146 and the Natural Sciences and Engineering Research Council (NSERC) of Canada under grant number SAPIN-2023-00029. 

\bibliography{ref}  

\end{document}